\documentclass[reprint,
 amsmath,amssymb,
pra,
]{revtex4-1}
\usepackage{dcolumn}% Align table columns on decimal point
\usepackage{bm}% bold math
\usepackage{amsfonts}
\usepackage{amsmath}
\usepackage{amssymb}
\usepackage{float}
\usepackage{graphicx}
\usepackage{subfigure}
\usepackage{epstopdf}
\usepackage{hyperref}

\begin{document}

\title{Optically induced phonon blockade in an optomechanical system with second-order nonlinearity}
\author{Hong Xie$^{1,2,3}$}
\author{Chang-Geng Liao$^{2,3,4}$}
\author{Xiao Shang$^{2,3}$}
\author{Zhi-Hua Chen$^{2,3}$}
\author{Xiu-Min Lin$^{2,3}$}
\thanks{xmlin@fjnu.edu.cn}

\affiliation{$^{1}$ College of JinShan, Fujian Agriculture and Forestry University, Fuzhou 350002, China}
\affiliation{$^{2}$ Fujian Provincial Key Laboratory of Quantum Manipulation and New Energy Materials, College of Physics and Energy, Fujian Normal University, Fuzhou 350117, China}
\affiliation{$^{3}$  Fujian Provincial Collaborative Innovation Center for Optoelectronic Semiconductors and Efficient Devices, Xiamen 361005, China}
\affiliation{$^{4}$ Department of Electronic Engineering, Fujian Polytechnic of Information Technology, Fuzhou, 350003, China}

\begin{abstract}
Quantum control of phonons has being become a focus of attention for developing quantum technologies. Here, we propose a proposal to realize phonon blockade in a quadratically coupled optomechanical system, where a strong nonlinear interaction between photons and phonons can be induced by an external field coherently driving the cavity, and the effective coupling strength is tunable by adjusting the amplitude of the driving field. This optically induced nonlinearity is different from standard methods for realization of phonon blockade, where the nonlinearity is achieved by coupling the mechanical system to superconducting qubits. We both analytically and numerically study the phonon statistical properties via the steady-state solution of the second-order correlation function, and find phonon blockade can be efficiently realized for a large cooperativity of the system, which is achievable based on the optically enhanced nonlinear coupling and high quality mechanical system.

\end{abstract}

\maketitle

\section{introduction}
Exploring the quantum feature of macroscopic objects is of great interest both for applications in the fields of quantum information and quantum sensing  \cite{Schwab2005Putting,Poot2012Mechanical} and for fundamental questions about quantum-classical boundary \cite{zurek1991decoherence}. The tremendous progress in the fabrication of mechanical device provides a wide range of mass and frequency mechanical systems \cite{aspelmeyer2014cavity}, which makes the experimental testing of macroscopic quantum phenomena possible. Particularly, with the advances of the optomechanical cooling technologies \cite{teufel2011sideband,chan2011laser,Clark2017Sideband}, many efforts have been paid to the quantum control of mechanical objects \cite{Hong2017Hanbury,Riedinger2016Non,o2010quantum}. For example, by coupling the mechanical mode to a superconducting qubit , several experiments have implemented the measurement on gigahertz phonons at the single phonon level \cite{o2010quantum,Reed2017Faithful,Chu2017Quantum}. A central goal of these experiments is to manipulate the mechanical degree of freedom in the true quantum regime, in analogy to controlling the quantum state of light in cavity \cite{Reiserer2015Cavity} or circuit QED systems \cite{Wang2016A}.

Phonon blockade is one of quantum features of mechanical oscillator \cite{liu2010qubit,didier2011detecting}, which comes from the original idea of photon blockade \cite{imamoglu1997strongly} in cavity QED systems. The blockade effect means that a single photon's (phonon's) appearing in the cavity (mechanical system) will prevent the exciting of the second one. To realize phonon blockade, the nonlinearity of mechanical system is required. Recently, the quantum nonlinearities have been proposed by coupling the mechanical mode to a qubit \cite{miranowicz2016tunable,wang2016method} or two-level defect \cite{ramos2013nonlinear}. The phonon blockade in the nonlinear mechanical system is mainly realized by utilizing the anharmonic energy-level \cite{liu2010qubit,didier2011detecting,miranowicz2016tunable,wang2016method,ramos2013nonlinear} or the destructive interference between different paths from ground state to two-phonon state \cite{xu2016phonon}. Although phonon blockade has been theoretically studied by combining mechanical oscillator with a two-level system, it is still a challenge for experimental demonstration compared with photon blockade in cavity \cite{Birnbaum2005Photon} or circuit QED systems \cite{Hoffman2011Dispersive}. The main obstacle is that the large nonlinearity for realization of phonon blockade requires the strong coupling limit, i.e., the coupling strength is much larger than the decay rates of mechanical oscillator and two-level system, which is difficult to realize under current experimental conditions.

In this letter, we study the phonon blockade in a quadratically coupled optomechanical system. The strong nonlinear coupling between photons and phonons can be effectively induced via an external field coherently driving the cavity, and the effective coupling strength is tunable by adjusting the amplitude of the driving field. An anharmonic energy-level of the system is obtained based on the nonlinear coupling, where the exciting of the second phonon is prevented due to large detuning when the first one has been excited by a weak field. The phonon statistical properties are studied with the second-order correlation function. The results show that strong phonon blockade can be observed for a large cooperativity of the system. It is worth stressing that the large nonlinearity in our system is achieved by the optical control of the mechanical system, which is different from previous studies where the nonlinearity is obtained through an ancillary two-level system \cite{liu2010qubit,didier2011detecting,miranowicz2016tunable,wang2016method,ramos2013nonlinear,xu2016phonon}.

The result is also different from our recent proposal \cite{Xie2017Phonon}, where the phonon blockade is realized based on $\chi^{(3)}$ Kerr-type nonlinearity in the quadratically coupled optomechanical system, while in this paper the phonon blockade is implemented by the effective $\chi^{(2)}$ nonlinearity. Normally, the $\chi^{(2)}$-based phonon blockade can potentially be implemented with a larger nonlinearity as compared to the case of kerr nonlinearity \cite{liu2010qubit,wang2016method}. Hence, the scheme proposed here is more achievable than Ref.~\cite{Xie2017Phonon}.

The paper is arranged as follows. In Sec. II, we introduce the theoretical model of the optomechanical system with quadratic coupling. Sec. III provides both
the analytical and numerical results of the correlation function, which is used to characterize the phonon statistical properties of the mechanical mode. Sec. IV presents the conclusions.

\section{theoretical model}
The quadratically coupled optomechanical system, where the cavity frequency is coupled to the square of the mechanical displacement, has been experimentally realized in a number of optomechanical devices, such as membrane-in-middle Fabry-P\'{e}rot cavities \cite{jayich2008dispersive,thompson2008strong,sankey2010strong}, microsphere-nanostring systems \cite{Brawley2014Nonlinear}, and photonic crystal optomechanical cavity \cite{paraiso2015position}. The quadratic coupling has also been proposed for the realization of mechanical non-classical states \cite{nunnenkamp2010cooling,asjad2014robust,tan2013dissipation,abdi2016dissipative,PhysRevA.95.053844}, optomechanically induced transparency \cite{huang2011electromagnetically,karuza2013optomechanically,zhan2013tunable} and opacity \cite{si2017optomechanically}, quantum nondemolition measurement of phonons, and single-photon nonlinearity \cite{liao2013photon,Liao2014Single,xie2016single}. The Hamiltonian of the system is written as ($\hbar=1$)
\begin{equation}
{H} = H_s+H_d+H_p,
\end{equation}
where
\begin{eqnarray}
{H_s} = {\omega _c}{a^\dag }a+{\omega _m}{b^\dag }b + {g_0}{a^\dag }a{(b + {b^\dag })^2}
\end{eqnarray}
is the Hamiltonian of the quadratically coupled optomechnical system without driving term, ${\omega _c}$ and $\omega _m$ are the frequency of cavity mode $a$ and mechanical mode $b$, ${g_0}$ denotes the single-photon quadratic coupling strength, and
\begin{eqnarray}
{H_d} = \Omega(a^{\dag}e^{-i\omega _Lt}+ae^{i\omega _Lt})
\end{eqnarray}
represents the coupling between cavity and the driving field with laser amplitude $\Omega$ and frequency $\omega _L$. The interaction between mechanical mode and the weak pumping field with amplitude $\varepsilon$ and frequency $\omega _p$ is described by
\begin{eqnarray}
{H_p} = \varepsilon(b^{\dag}e^{-i\omega _pt}+be^{i\omega _pt}),
\end{eqnarray}
which is used to excite phonons in the mechanical mode. In general, the weak pumping field for the mechanical mode can be implemented by an dc voltage source \cite{szorkovszky2013strong} or piezoelectric radio signal \cite{fan2015cascaded}.

In the rotating picture at the driving frequency ${\omega _L}$, the Hamiltonian of the system can be rewritten as
\begin{equation}
{H_1} = {\Delta _c}{a^\dag }a+{\omega _m}{b^\dag }b + {g_0}{a^\dag }a{(b + {b^\dag })^2}+\Omega(a^{\dag}+a)+H_p
\end{equation}
with the detuning  ${\Delta _c} = {\omega _c} - {\omega _L}$ between the cavity mode and driving field. By means of the standard linearizaion procedure of cavity optomechanics, we split both the cavity and mechanical modes into an average amplitude and a fluctuation term, i.e., $ a \rightarrow\alpha  + a $ and $ b \rightarrow \beta  + b$. The average coherent amplitudes can be obtained as $\alpha = {\Omega}/({-\Delta _c+i\kappa/2})$ and $\beta=0$ based on the Heisenberg-Langevin equations in the steady-state case \cite{tan2013dissipation,xie2016single}, where $\kappa$ is cavity decay rate. Note that $\alpha$ can be chosen real by a proper choice of the phase of driving field.

After the standard linearizaion procedure, the Hamiltonian can be expressed as
\begin{equation}
  {H_2} = {\Delta _c}{a^\dag }a + {\omega_m}{b^\dag }b +{g}(a^\dag +a){(b + {b^\dag })^2}+H_p,
\end{equation}
where $g=g_0\alpha$ is the effective coupling strength which can be tuned by controlling the amplitude of the driving field. The small second-order term ${g_0}{a^\dag }a{(b + {b^\dag })^2}$ has been neglected in comparison to ${g}(a^\dag +a){(b + {b^\dag })^2}$. When the cavity is driven on the red two-phonon resonance, i.e., ${\Delta _c}=2{\omega_m}$, the rapid oscillation terms with high frequencies $ \pm ({\Delta _c} + 2{\omega _m})$ and $\pm {\Delta _c}$ can be neglected under the rotating-wave approximation in the case of ${g}\ll {\omega _m}$. Thus the Hamiltonian becomes
\begin{eqnarray}
  {H_3} = {\Delta _c}{a^\dag }a+ {\omega _m}{b^\dag }b + g({a^\dag }{b^2} + a{b^\dag }^2)+H_p.
\end{eqnarray}
Applying the unitary transformation
\begin{eqnarray}
  U(t) = \exp{(i2{a^\dag }a\omega_pt+i{b^\dag }b\omega_pt)}
\end{eqnarray}
to the Hamiltonian ${H_3}$ makes the pump term time-independent, and a new Hamiltonian $H_{\text{eff}}=UH_3U^\dag-i\partial U/\partial t$ is generated with the form
\begin{equation}\label{eff}
{H_\text{eff}} = {(\Delta _c-2\omega_p)}{a^\dag }a+ {\Delta _p}{b^\dag }b + g({a^\dag }{b^2} + a{b^\dag }^2)+\varepsilon(b^{\dag}+b),
\end{equation}
where $\Delta _p=\omega _m-\omega_p$ is the detuning between mechanical mode and pump field, and $\Delta _c-2\omega_p=2\Delta _p$ satisfies the two-phonon resonant condition in the rotating frame corresponding to the unitary operator $U(t)$. We note that the effective Hamiltonian $H_\text{eff}$ has been used to study cooling and squeezing \cite{nunnenkamp2010cooling}, photon blockade \cite{xie2016single}, and macroscopic quantum superposition of mechanical mode \cite{tan2013dissipation}.
Especially, the Hamiltonian $H_\text{eff}$ presents an effective $\chi^{(2)}$ nonlinearity, where the creation (annihilation) of a photon is accompanied by the annihilation (creation) of two phonons. This enable a resonant interaction between states $\left| {{n_a},{m_b}} \right\rangle $ and $\left| {({n} - 1)_a,({m} + 2)_b} \right\rangle$, where ${{n_a}}$ (${m_b}$) denotes the photon (phonon) number of the cavity (mechanical) mode.

\begin{figure}[!htb]
  {\includegraphics[width=0.4\textwidth]{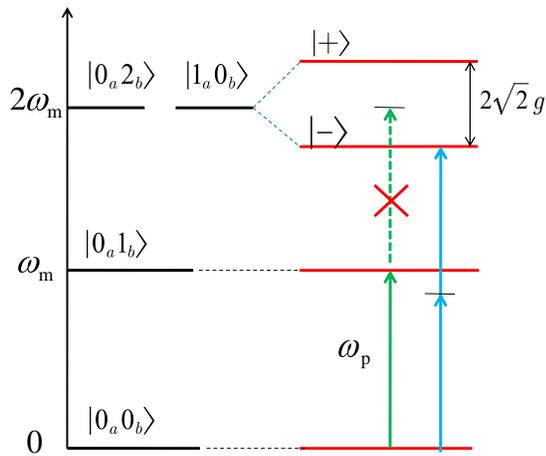}}
\caption{(Color online) Anharmonic energy-level diagram of the system in the few-phonon subspace. States are label as $|n_a,m_b\rangle$, where $n_a$ ($m_b$) denotes the photon (phonon) number. The nonlinear interaction between photon and phonons splits the degeneracy of the two states in the two-phonon subspace. Green arrows present the single-phonon transition; blue arrows, the two-phonon transition.}
\end{figure}
To see this more clearly, the Hamiltonian ${H_\text{eff}}$ without the pump term is diagonalized in the two-phonon subspace. The eigenstates of the system are given by
\begin{eqnarray}
  \left| {{\pm }} \right\rangle  = \frac{1}{{\sqrt 2 }}(\left| {0_a2_b} \right\rangle  \pm \left|{1_a0_b}  \right\rangle)
\end{eqnarray}
with the corresponding eigenvalues $\pm\sqrt{2}g$.

The energy level of the system in the few-phonon subspace is shown in Fig.~1. Note that the degeneracy of the two states in the two-phonon subspace is split due to nonlinear coupling between photon and phonons. In analogous to Jaynes-Cummings ladder of the atom-cavity system for implementing photon blockade \cite{imamoglu1997strongly}, the anharmonic level of the driven quadratically coupled optomechanical system is the crucial feature to realize phonon blockade.

After being taken into account the cavity and mechanical dissipation, the dynamical evolution of the system is described by the master equation
\begin{eqnarray}\label{master}
  \dot \rho  = &-i[H_\text{eff},\rho ] + \frac{\kappa}{2}(2a\rho {a^\dag}-{a^\dag }a\rho  - \rho {a^\dag }a) \nonumber \\ &
   + \frac{\gamma}{2}{\bar n}_{th}(2b^\dag\rho {b}-{b}b^\dag \rho  - \rho {b}b^\dag) \nonumber \\ &
   + \frac{\gamma}{2}({\bar n}_{th}+1)(2b\rho {b^\dag}-{b^\dag }b\rho  - \rho {b^\dag }b),
\end{eqnarray}
where $\gamma$ represents the mechanical decay rate, and ${\bar n}_{th}=1/[\exp{(\hbar {\omega _m}/{k_B}T)}-1]$ denotes the thermal phonon number at the environmental temperature $T$.

\section{phonon blockade}
\subsection{Steady-state solution of correlation function}
To characterize the phonon statistical properties of the mechanical mode, we study the steady-state second-order correlation function
\begin{eqnarray}\label{g2}
   g^{(2)}(0)=\frac{\langle{b^{\dag}b^{\dag}bb}\rangle}{{\langle{b^{\dag}b}\rangle}^2}=\frac{tr({\rho_{ss}b^{\dag}b^{\dag}bb})}{tr({\rho_{ss}b^\dag}b)^2},
\end{eqnarray}
where $\rho_{ss}$ is the density matrix of the system in the steady-state case. The second-order correlation function in quantum optics domain is originally defined for measuring the quantum nature of optical fields, which is also suitable for all bosonic fields. Recently, the second-order correlation function of phonon field in cavity optomechanics has been measured by Hanbury-Brown-Twiss type experiment \cite{Hong2017Hanbury}, which is used to verify a single-phonon exciting in an optomechanical resonator. The equal-time second-order correlation function ${g^{(2)}}(0)<1$ (${g^{(2)}}(0)>1$) represents phonon antibunching (bunching), and the limit ${g^{(2)}}(0) \to 0$ corresponds to the complete phonon blockade regime in which only one phonon can be excited to the mechanical mode.
\begin{figure}[t]
 {\includegraphics[width=0.48\textwidth]{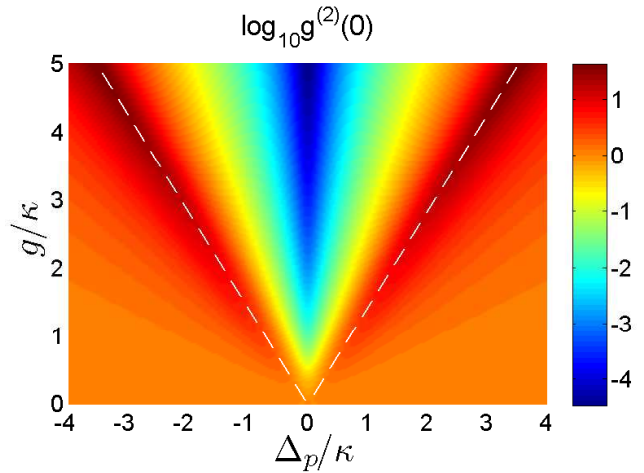}}
\caption{(Color online) Logarithmic plot of second-order correlation function as a function of $\Delta_p/\kappa$ and $g/\kappa$. The white dash curve is plotted with $\Delta_p=\pm\sqrt{2}g/2$. Parameters are chosen as $\gamma/\kappa=0.01$, $\varepsilon/\kappa=0.1$, and $\overline{n}_{th}=0$.}
\end{figure}

Figure 2 shows the dependence of the logarithmic second-order correlation function on the detuning $\Delta_p/\kappa$ and the effective coupling strength $g/\kappa$. The phonon antibunching ${g^{(2)}}(0)<1$ appears at detuning $\Delta_p=0$. This can be understood from the anharmonic energy-level of the system, where if a single-phonon transition $|0_a0_b\rangle\rightarrow |0_a1_b\rangle$ happens, the subsequent transition $|0_a1_b\rangle\rightarrow |\pm\rangle$ is suppressed for large detuning, as shown with green arrows in Fig.~1. In this case, the first excited phonon in the mechanical mode will prevent the second one being excited.  At detuning $\Delta_p=\pm\sqrt{2}g/2$, phonon bunching ${g^{(2)}}(0)>1$ can be observed in the strong coupling regime. This bunching comes from the two-phonon transition $|0_a0_b\rangle\rightarrow |\pm \rangle$, as shown with the blue arrows in Fig.~1. The phonon bunching means that the mechanical mode tends to be occupied by two phonons rather single one. Figure 2 also shows that both the phonon antibunching and bunching become stronger with the increasing of coupling strength $g/\kappa$.

\subsection{Analytical solution}

To obtain detailed conditions for realizing the above discussed phonon antibunching and bunching, we approximately calculate the correlation function in a truncated Fock state basis.

Assume that the system is initially prepared in its ground state. For a sufficient weak pumping field, only few phonons can be excited in the mechanical mode. Thus the general wave function of the system can be expanded in the few-phonon subspace as
 \begin{equation}\label{span}
  \left|\psi\textrm{(}t\textrm{)}\right> =C_{00}\left|0_a0_b\right>+C_{01}\left|0_a1_b\right>+C_{02}\left|0_a2_b\right>+C_{10}\left|1_a0_b\right>,
 \end{equation}
where $C_{n,m}$ is the probability amplitude, and $|C_{n,m}|^2$ denotes occupying probability in the state $\left|n_am_b\right>$. To check the validity of the state truncation in the few-phonon subspace, the fidelity is defined as the sum of occupying probabilities of four states in Eq.~(\ref{span})
 \begin{equation}\label{F}
 F =|C_{00}|^2+|C_{01}|^2+|C_{02}|^2+|C_{10}|^2.
 \end{equation}
When the fidelity satisfies $F\approx1$, the state expansion in Eq.~(\ref{span}) is valid. The steady-state solution of the fidelity based on master equation (\ref{master}) is shown in Fig. 3. For a weak pumping field $\varepsilon<0.5\kappa$, the fidelity is almost equal to $1$, and decreases with the  increasing pumping strength $\varepsilon$. If the fidelity $F$ is smaller than 1, the multi-phonon states will be excited. In this case, it is not enough to expand the state in the above four-state basis space.

\begin{figure}[t]
 {\includegraphics[width=0.4\textwidth]{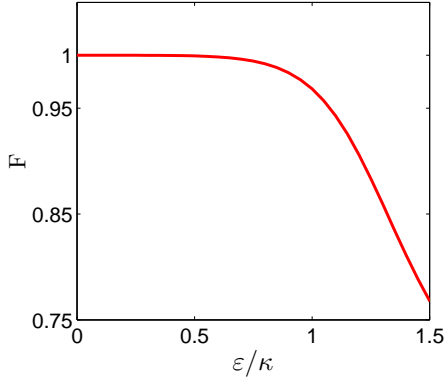}}
\caption{(Color online) The fidelity F given by Eq.~(\ref{F}) as a function of $\varepsilon/\kappa$. Parameters are chosen as $g/\kappa=2$, $\gamma/\kappa=0.01$, $\Delta_p=0$, and $\overline{n}_{th}=0$.}
\end{figure}

When the dissipations of cavity and mechanical modes are included to calculate the approximate solution of the probability amplitudes, an effective non-Hermitian Hamiltonian is considered
\begin{eqnarray}
 H'_\text{eff} = H_\text{eff} - i\frac{\kappa }{2}{a^\dag }a - i\frac{\gamma }{2}{b^\dag }b.
\end{eqnarray}
Based on the Schr\"{o}dinger equation $i{d\left| \psi  \right\rangle }/{{dt}} = H'_\text{eff}\left| \psi  \right\rangle$, the motion equations for the probability amplitudes are given by
\begin{subequations}\label{test}
\begin{align}
 &\dot{C}_{00}=-i\varepsilon  C_{01},
 \\
 &\dot{C}_{01}=-i\varepsilon  C_{00}-i(\Delta_p-i\frac{\gamma}{2})C_{01}-i\sqrt{2}\varepsilon C_{02},
\\
  &\dot{C}_{02}=i\sqrt{2}\varepsilon C_{01}-i(2\Delta_p-i\gamma)C_{02}-i\sqrt{2}g C_{10},
\\
  &\dot{C}_{10}=-i(2\Delta_p-i\frac{\kappa}{2}))C_{10}-i\sqrt{2}g C_{02}.
\end{align}
\end{subequations}
In the weak pumping regime, $C_{01}$ is proportional to $\varepsilon$ while $C_{02}$ and $C_{10}$ are proportional to $\varepsilon^2$. In the limit $\varepsilon\rightarrow0$, $C_{00}\approx1$, i.e., the system remains in the ground state. By neglecting the higher order terms of $\varepsilon$, the steady-state solution of the probability amplitudes can be approximately given by
\begin{subequations}\label{amp}
\begin{align}
 &{C}_{01}=\frac{-i\varepsilon}{\gamma/2+i\Delta_p},
 \\
 &{C}_{02}=\frac{-\sqrt{2}\varepsilon^2(\kappa/2+i2\Delta_p)}{[(\gamma+i2\Delta_p)(\kappa/2+i2\Delta_p)+2g^2](\gamma/2+i\Delta_p)},
 \\
 &{C}_{10}=\frac{-i2g\varepsilon^2}{[(\gamma+i2\Delta_p)(\kappa/2+i2\Delta_p)+2g^2](\gamma/2+i\Delta_p)}.
\end{align}
\end{subequations}

\begin{figure}[!t]
  {\includegraphics[width=0.4\textwidth]{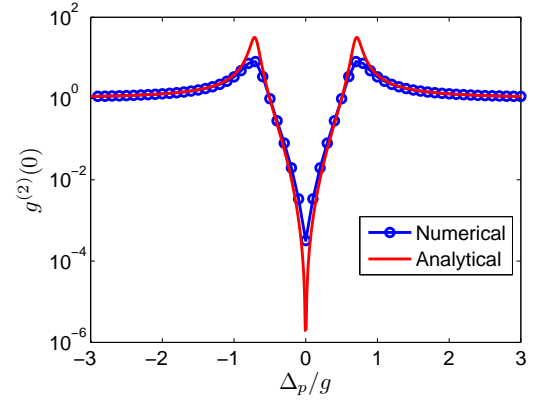}}
  \caption{(Color online) Second-order correlation function ${g^{(2)}}(0)$ as a function of $\Delta_p$. Parameters are taken as $g/\kappa=2$, $\gamma/\kappa=0.01$, $\varepsilon/\kappa=0.1$, and $\overline{n}_{th}=0$.}
\end{figure}

For the state given by Eq.~(\ref{span}), the correlation function can be written as
\begin{eqnarray}\label{g21}
   g^{(2)}(0)=\frac{2|C_{02}|^2}{(|C_{01}|^2+2|C_{02}|^2)^2}\approx\frac{2|C_{02}|^2}{|C_{01}|^4},
\end{eqnarray}
where we have used the relation $|C_{02}|\ll|C_{01}|$ to obtain the second approximated expression. Based on probability amplitudes in Eq.~(\ref{amp}), we have
\begin{eqnarray}\label{g22}
   g^{(2)}(0)=\frac{|(\gamma/2+i\Delta_p)(\kappa/2+i2\Delta_p)|^2}{|(\gamma/2+i\Delta_p)(\kappa/2+i2\Delta_p)+g^2|^2}.
\end{eqnarray}
When the pumping field is tuned to resonance transition frequency of the mechanical mode, i.e., $\Delta_p=0$, the correlation function reduces to
 \begin{eqnarray}\label{g23}
   g^{(2)}(0)=\frac{1}{(1+4g^2/\kappa\gamma)^2}.
\end{eqnarray}
In this case, phonon antibunching ${g^{(2)}}(0)< 1$ always exists in the present of optomechanical coupling. The correlation function ${g^{(2)}}(0)=1$ if $g=0$, which is a feature of coherent state. Obviously, the phonon blockade ${g^{(2)}}(0) \ll 1$ can be realized when the cooperativity $C$ satisfies
\begin{equation}\label{C}
  C=\frac{4g^2}{\kappa\gamma}\gg1.
\end{equation}
This result is similar with that in the Jaynes-Cummings model, where a large cooperativity $C\gg1$ is required for the observation of strong photon antibunching \cite{Birnbaum2005Photon}.

When the pumping field is tuned to two-phonon resonance transition, i.e., $\Delta_p=\pm\sqrt{2}g/2$, the correlation function becomes
 \begin{equation}\label{g24}
   g^{(2)}(0)=\frac{(\gamma^2/4+g^2/2)(\kappa^2/4+2g^2)}{(\gamma^2/4+g^2/2)(\kappa^2/4+2g^2)+g^2\kappa\gamma/2-g^4}.
\end{equation}
The phonon bunching ${g^{(2)}}(0)> 1$ can be observed if $g^4>g^2\kappa\gamma/2$, i.e., $C/2>1$.

\begin{figure}[!t]
  {\includegraphics[width=0.45\textwidth]{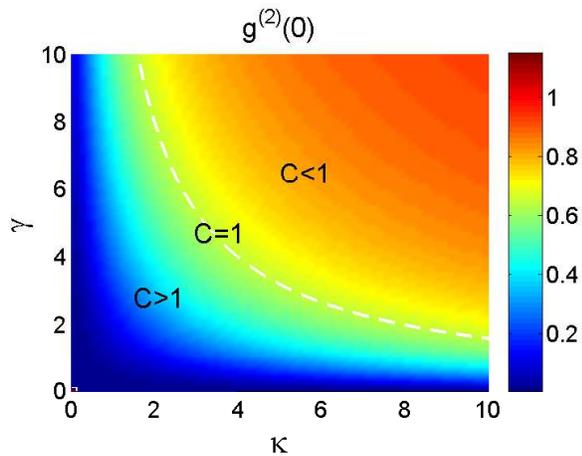}}
  \caption{(Color online) Second-order correlation function ${g^{(2)}}(0)$ as a function of $\kappa$ and $\gamma$ at $\Delta_p=0$. Parameters are taken as $g=2$, and $\varepsilon=0.1$, and $\overline{n}_{th}=0$.}
\end{figure}

\subsection{Phonon blockade}
In order to comprehend the analytical result more clearly, the second-order correlation function ${g^{(2)}}(0)$ as a function of $\Delta_p/g$ is plotted in Fig.~4. The red curve denotes the analytical solutions of Eq.~(\ref{g22}), while the blue curve with circles is a snapshot taken from Fig.~2, which is the numerical result based on master equation Eq.~(\ref{master}). The two curves are well consistent. The dip ${g^{(2)}}(0) \ll 1$ at $\Delta_p=0$ exactly confirms phonon blockade. As discussed above, the phonon blockade comes from the suppression of the transitions $|0_a1_b\rangle\rightarrow |\pm\rangle$ due to large detuning. It should be noted that the detuning must be greater than the mechanical decay rate, i.e., $\sqrt{2}g\gg\gamma$, so that the transition $|0_a0_b\rangle\rightarrow |0_a1_b\rangle$ can be spectrally resolved, and strong phonon blockade is realized.

Figure 5 shows the correlation function ${g^{(2)}}(0)$ as a function of $\kappa$ and $\gamma$ at $\Delta_p=0$. The white curve is plotted by $C=1$. It is clear that phonon antibunching still presents even in the case of the large cavity and mechanical decay rates $\kappa,\gamma>g$. However, strong phonon blockade can only be observed in the weak decay rate regime $\kappa,\gamma\ll g$, i.e., the cooperativity $C \gg 1$. These results are consistent with the analytical solution of Eq.~(\ref{g23}).

\begin{figure}[!t]
  {\includegraphics[width=0.42\textwidth]{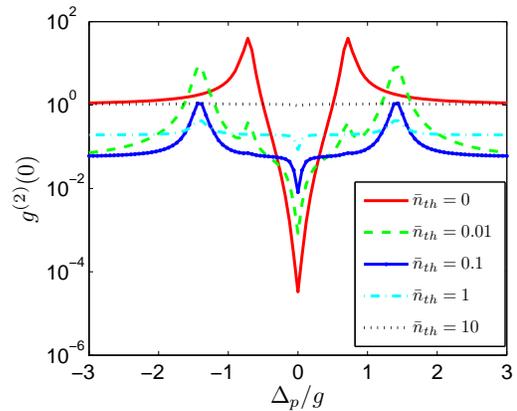}}
  \caption{(Color online) Second-order correlation function ${g^{(2)}}(0)$ as a function of cavity decay rate $\Delta_p$ at different value of $\overline{n}_{th}$.  Parameters are the same as Fig.~4.}
\end{figure}
So far we have only considered the case where the mechanical mode is initially cooled to its ground state, and the mechanical thermal mean phonon number is chosen as $\bar{n}_{th}=0$. Next, we consider a finite but small equilibrium occupation number of the mechanical mode.
Figure 6 displays ${g^{(2)}}(0)$ as a function of $\Delta_p/g$ at different thermal mean phonon number $\bar{n}_{th}$. At detuning $\Delta_p=0$, we find the phonon blockade survives for a small $\bar{n}_{th}$, but its quality is degraded with the increasing of $\bar{n}_{th}$. This is because the single-phonon excitation will be destroyed by thermal noise when the thermal mean phonon number $\bar{n}_{th}$ approaches to the average phonon number in the mechanical mode. So in order to improve quality of phonon blockade, the thermal mean occupation number should be sufficient small.

Figure 6 also reveals that with the increasing of $\bar{n}_{th}$, the phonon bunching at $\Delta_p=\pm\sqrt{2}g/2$ disappears, but new bunching appears at $\Delta_p=\pm\sqrt{2}g$. Qualitatively, the new bunching emerges from the transitions $|0_a1_b\rangle \rightarrow |\pm\rangle$. When the thermal mean phonon number $\bar{n}_{th}$ increases (but is still less than 1), the probability of occupation in one-phonon state $|0_a1_b\rangle$ is enhanced. Consequently, the probability of the transitions from $|0_a1_b\rangle$ to $|\pm\rangle$ is also raised. However, for a thermal mean phonon number $\bar{n}_{th}>1$, both phonon antibunching and bunching are destroyed by thermal noise.

\section{Discussion and Conclusions}
In conclusion, we have realized the phonon blockade in a optomechanics system with quadratic coupling, where the nonlinearity of mechanical resonator is induced by driving the cavity on two-phonon red sideband. For the system being initially cooled to its ground state, only a single phonon can be excited in the mechanical mode due to the nonlinearity of the mechanical system.

By calculating the second-order correlation function both numerically and analytically in the steady-state case, we have found the conditions for the realization of strong phonon blockade. First, a large cooperativity of the system is required $C\gg1$, i.e., $g\gg\kappa,\gamma$. Second, the thermal excitation number should be small enough, so that the single-phonon excitation in the mechanical mode will not be destroyed by the thermal noise.

In recent experiment for quadratically coupled optomechanical system, the single-photon quadratic coupling strength $g_0$ is about 245 Hz in a planar silicon photonic crystal cavity \cite{paraiso2015position}, and is likely enhanced to 1 kHz \cite{kalaee2016design} or 100 kHz \cite{paraiso2015position} by carefully tuning the photonic crystal structure. Thus the effective coupling strength $g=g_0\alpha$ up to hundreds of MHz can be obtained for $\alpha\sim10^4$ \cite{sankey2010strong}. With a high mechanical quality factor $Q\sim10^7$ ($\omega_m\approx5.6$ GHz and $\gamma\approx328$ Hz) \cite{meenehan2015pulsed} and a cavity decay rate about 20 MHz \cite{sekoguchi2014photonic} in the photonic crystal cavity, the strong coupling condition $g\gg\kappa,\gamma$ is possibly implemented with current experimental parameters. In addition, for a mechanical system with gigahertz frequency at cryogenic temperature ($25$ mK) \cite{o2010quantum,Chu2017Quantum,Hong2017Hanbury}, a small thermal mean phonon number $\bar{n}_{th}\sim 10^{-5}$ can be achieved. All of these make it possible to realize phonon blockade under current experimental conditions.

\section*{Acknowledgments}
 We acknowledge supports from the National Natural Science Foundation of China (Grants No.~61275215 and No.~11674059), the Natural Science Foundation of Fujian Province of China (Grants No.~2016J01008 and No.~2016J01009), the Educational Committee of Fujian Province of China (Grants No.~JAT160687, No.~JA14397 and No.~JAT170148), and the College Funds for Distinguished Young Scientists of Fujian Province of China.

\bibliography{bibfile}
\bibliographystyle{apsrev4-1}

\end{document}